\documentclass{article}
\PassOptionsToPackage{authoryear}{natbib}  

\usepackage[margin=1in]{geometry}

\usepackage{amsmath,amssymb,amsfonts,amsthm}

\usepackage{graphicx}
\usepackage{booktabs}
\usepackage{multirow}

\usepackage[numbers]{natbib}
\usepackage[hidelinks]{hyperref}

\usepackage{microtype}
\usepackage{amsmath, amssymb}
\usepackage{graphicx}
\usepackage{placeins}
\setkeys{Gin}{draft=false}
\usepackage{booktabs}
\usepackage{xcolor}
\usepackage{url}
\usepackage{wrapfig}
\usepackage{float}
\usepackage{subcaption}
\usepackage{xcolor}
\usepackage{colortbl}
\usepackage{comment}
\usepackage[normalem]{ulem}

\definecolor{mikepurple}{RGB}{128, 70, 170}

\newcounter{ToDo}
\newcounter{gaocomm} 
\newcounter{Note}
\definecolor{blue-violet}{rgb}{0.00,0.75,0.90}
\definecolor{mygreen}{rgb}{0.0, 0.5, 0.0}
\definecolor{awesome}{rgb}{1.0, 0.13, 0.32}
\definecolor{bostonuniversityred}{rgb}{1.0, 0.0, 0.0}

\title{Structured SDEs for Realized GARCH: Continuous-Time Volatility Modeling with Component-Level Analysis}

\author{
    Hongyu Yao \qquad
    Zhengyang Chi \qquad
    Junbin Gao \\[0.8em]
    The University of Sydney \\
    Sydney, NSW, Australia \\[0.5em]
    \texttt{hyao8085@uni.sydney.edu.au} \\
    \texttt{zchi7292@uni.sydney.edu.au} \\
    \texttt{junbin.gao@sydney.edu.au}
}

\date{}

\begin{document}
\raggedbottom
\maketitle

\begin{abstract}
We propose a structured continuous-time stochastic volatility model for one-day-ahead integrated-variance forecasting. The model introduces a jointly evolving realized-measure state that feeds directly into latent variance dynamics, while correlated Brownian innovations capture return-linked dependence. Parameters are estimated through simulation-based predictive learning, and Monte Carlo simulation forecasts integrated variance. Across 17 equity indices, the proposed model shows strong out-of-sample performance relative to GARCH, GARCH-It\^{o}, Realized GARCH, HAR, and HEAVY benchmarks under QLIKE, MSE, and MAE. Ablation results show that realized-measure feedback provides substantial incremental forecasting value within the proposed framework, while return-linked Brownian dependence contributes more modest complementary gains. The results support incorporating measurement-driven information directly into continuous-time latent volatility dynamics to improve integrated-variance forecasting.
\end{abstract}

\section{Introduction}
\label{introduction}
Volatility forecasts characterize the variability of future asset returns and are critical for risk management, derivative pricing, and portfolio allocation~\citep{Cont2001,HansenLunde2005,Shephard2023}. A latent instantaneous variance process $s_t$ governs the local intensity of the return variation at each point in time, while volatility over a finite horizon depends on the cumulative variation generated along the path of $s_t$. Since $s_t$ is unobserved and time-varying, volatility forecasts require modeling the evolution of this latent process~\citep{Heston1993,Shephard2023}. For next-day-ahead forecasting, the relevant target is the integrated variance $H_{n+1}$, which accumulates the instantaneous variance on the next trading day~\citep{Andersen2001}.

\begin{equation}
H_{n+1}
:=
\int_n^{n+1} s_t\,dt.
\label{eq:integrated_variance}
\end{equation}

Different volatility models encode the latent variance in different ways. Continuous-time stochastic volatility models describe variance evolution through stochastic differential equations (SDEs) and can represent the dependence of return-volatility through correlated Brownian motions~\citep{Heston1993,Masoliver2002, Chen2024}. In contrast, GARCH-type models update conditional variance discretely using observed returns~\citep{Bollerslev1986}. High-frequency realized measures provide additional information about daily integrated volatility~\citep{Andersen2001,BarndorffNielsen2002,Song2021},   motivating measurement-driven models such as Realized GARCH and HEAVY~\citep{Hansen2012,Shephard2010}. 

These measurement-driven models incorporate realized measures through discrete-time updates. Existing continuous-time volatility models capture intraperiod variance evolution, but they typically do not incorporate a jointly evolving realized-measure feedback mechanism in the latent variance dynamics. This leaves a gap in how measurement-driven information is incorporated into continuous-time stochastic volatility models and raises the question of whether such feedback provides additional forecasting value when embedded in a continuous-time stochastic system.

Inspired by the measurement-driven structure of Realized GARCH~\citep{Hansen2012}, we address this gap by proposing a structured continuous-time stochastic volatility model in which the information in the realized-measure state feeds directly into the latent variance dynamics. We further allow return innovations to be correlated with innovations in both the latent-variance and realized-measure states.

Our main research question is: \emph{Do realized-measure feedback and continuous-time stochastic dependence improve one-day-ahead integrated-variance forecasting?} We address this question by comparing the proposed model with established volatility benchmarks and by conducting controlled ablation studies that separately assess the contributions of realized-measure feedback and stochastic dependence.

\paragraph{Contributions.}
We make three main contributions.
First, we develop a continuous-time volatility model in which realized-measure information enters directly into the dynamics of latent variance state.
Second, we use correlated Brownian innovations to represent the dependence of returns and both latent-variance and realized-measure shocks, providing a parsimonious and economically interpretable dependence structure without imposing an additional parametric leverage function.
Third, we provide evidence of improved integrated variance forecast accuracy across equity markets. The gains are primarily associated with realized-measure feedback, while correlated Brownian dependence provides an additional improvement in predictive performance.

\section{Methodology}
\subsection{Structured Continuous-Time State System}
\label{Structured Continuous-Time State System}
Let $Y_t$ denote the log-price process, and let $x_t$ and $z_t$ denote unconstrained states associated with the instantaneous latent variance $s_t$ and the continuous-time realized-measure state $R_t$, respectively. We define $s_t = e^{x_t}$ and $R_t = e^{z_t}$ so that both variance-related quantities remain positive. 

The proposed framework introduces a realized-measure feedback function $g(\cdot)$ to govern how the realized-measure state $z_t$ affects the latent variance dynamics.  The proposed system is defined as follows:
\begin{align}
dY_t
&=
\sqrt{s_t}\,dW_t^Y,
\label{eq:new_1}
\\
dx_t
&=
\kappa_x(\theta-x_t)\,dt
+
\gamma g(z_t)\,dt
+
\eta_x\,dW_t^x,
\label{eq:new_2}
\\
dz_t
&=
\kappa_z(\xi+\phi x_t-z_t)\,dt
+
\eta_z\,dW_t^z
+
\sigma_u\,dU_t,
\label{eq:new_3}
\end{align}
The processes $W_t^Y$, $W_t^x$, and $W_t^z$ are Brownian motions driving the return, latent log-variance, and log realized-measure states, respectively, while $U_t$ is an independent Brownian motion representing additional measurement-specific variation in $z_t$. The parameters $\kappa_x$, $\theta$, $\gamma$, $\kappa_z$, $\xi$, $\phi$, $\eta_x$, $\eta_z$, and $\sigma_u$ are learnable, with their roles described in the following paragraphs. 

\paragraph{Log price process.}
During a trading interval, increments of log prices $Y_t$ correspond to the observed log returns $r_{t, t+\Delta}=Y_{t+\Delta}-Y_t$. This equation follows the standard continuous-time stochastic-volatility representation, in which the diffusion coefficient determines its conditional variance~\citep{Heston1993}. In particular, 
$\operatorname{Var}\!\left(dY_t \mid \mathcal{F}_t\right)
=s_t\,dt$ and $\operatorname{Var}(dW_t^Y)=dt$. Thus, the return diffusion is scaled by $\sqrt{s_t}$~\citep{Heston1993, Takahashi2023} in Eq.~\eqref{eq:new_1}.

\paragraph{Latent variance dynamics.}
The latent log-variance state in Eq.~\eqref{eq:new_2} follows a mean-reverting diffusion augmented with realized-measurement feedback. $\kappa_x(\theta-x_t)$ represents mean reversion part and captures the persistence in volatility~\citep{Cox1975, barndorff-nielsen_integrated_2003}. Here, $\theta$ is the long-term mean of log variance, and $\kappa_x$ controls the speed at which $x_t$ is pulled back towards that level.

The feedback function $g(\cdot)$ allows realized-measure information to enter the drift of the latent variance process. In the current specification, we adopt the identity map $g(u)=u$ for interpretability. Thus, the feedback term reduces to $\gamma z_t$. The coefficient $\gamma$ controls the strength of this feedback. This mechanism is similar to the measurement-driven role of realized information in Realized GARCH~\citep{Hansen2012, Song2021}. Appendix~\ref {Appendix:B} provides further details on the conceptual connection between the proposed model and Realized GARCH.

The diffusion term $\eta_x\,dW_t^x$ captures unpredictable innovations in latent volatility~\citep{Heston1993}. The parameter $\eta_x$ controls the magnitude of these innovations and can be interpreted as a volatility-of-volatility parameter.

\paragraph{Realized-measure dynamics.}
At each daily observation time $n$, the realized-measure state is anchored to the observed realized variance (RV), $R_n = RV_n$, and $z_n = \log RV_n$. Between observation times, $z_t$ evolves according to Eq~\eqref{eq:new_3}. This dynamic is motivated by the role of RV as an informative observable proxy for integrated variance, while allowing for measurement variation around the underlying volatility process~\citep{Patton2011}. We therefore model the conditional mean-reversion level of the log realized-measure state $z_t$ as $\xi+\phi x_t$, where $\xi$ determines its baseline level and $\phi$ controls its response to the latent log variance $x_t$. We impose $\phi>0$ so that a higher latent variance raises the mean-reversion level of the realized-measure state, consistent with its interpretation as a proxy for latent volatility~\citep{Hansen2012}.

The diffusion term $\eta_z\,dW_t^z$ allows the realized-measure state to evolve stochastically around the conditional mean-reversion level~\citep{Takahashi2023}.
This captures innovations in RV that are not explained by the current latent variance state~\citep{ShirotaHizuOmori2014, BekiermanGribisch2016}. 
We also include a separate term, $\sigma_u\,dU_t$, to represent an independent measurement-specific variation\citep{Hansen2012}. This parsimonious specification captures the relationship between latent volatility and RV while allowing additional measurement variation through the diffusion terms.

\paragraph{Existence and uniqueness.}
Let $Z_t=(x_t,z_t)^\top$ denote the transformed state process. Under the main specification, $g(u)=u$, the state dynamics can be written jointly as:
\begin{equation}
    dZ_t
    =
    (AZ_t+c)\,dt
    +
    B\,d\widetilde W_t,
\end{equation}
where
\begin{equation}
    A
    =
    \begin{pmatrix}
        -\kappa_x & \gamma\\
        \kappa_z \phi & -\kappa_z
    \end{pmatrix},\,
    c
    =
    \begin{pmatrix}
        \kappa_x\theta\\
        \kappa_z \xi
    \end{pmatrix},\,
    B
    =
    \begin{pmatrix}
        \eta_x & 0 & 0\\
        0 & \eta_z & \sigma_u
    \end{pmatrix},\,
    \text{and }
    d\widetilde W_t
    =
    \begin{pmatrix}
    dW_t^x\\
    dW_t^z\\
    dU_t
    \end{pmatrix}
    \label{eq:joint_dynamic_matrix}
\end{equation}

Here, $d\widetilde W_t$ collects the Brownian drivers of the state dynamics, with their dependence structure specified separately. The affine drift and constant diffusion coefficients are globally Lipschitz and satisfy the standard linear-growth condition. Hence, for any finite initial state, standard existence and uniqueness results for SDEs imply that it admits a unique strong solution on every finite time interval~\citep{Oksendal2003, KloedenPlaten1992}.

Given a solution for $x_t$, the log-price process is defined by
\begin{equation}
    Y_t
    =
    Y_0
    +
    \int_0^t e^{x_u/2}\,dW_u^Y.
\end{equation}
Because $x_t$ has continuous sample paths, it is almost surely bounded on every compact interval $[0,T]$. Consequently,
\begin{equation}
    \int_0^T e^{x_t}\,dt
    <
    \infty
    \qquad \text{a.s.}
\end{equation}
which ensures that the It\^{o} integral above is well defined over any finite forecasting horizon~\citep{Oksendal2003}. Finally, the transformations
\[
    s_t=e^{x_t},
    \qquad
    R_t=e^{z_t},
\]
preserve strict positivity of the instantaneous variance and realized-measure states by construction.

\paragraph{Sufficient stability condition.}
\label{Sufficient Stability Condition}

For the deterministic state dynamics, asymptotic stability requires the drift matrix $A$ defined in Equation~\eqref{eq:joint_dynamic_matrix} to be Hurwitz~\citep{Khalil2002}. Since
\[
    \operatorname{tr}(A)=-(\kappa_x+\kappa_z)<0,
    \qquad
    \det(A)=\kappa_z(\kappa_x-\gamma\phi)>0,
\]
a sufficient stability condition is
\[
    \kappa_x>0,
    \qquad
    \kappa_z>0,
    \qquad
    \gamma\phi<\kappa_x.
    \label{eq:sufficient_stability}
\]

The restriction is consistent with the modeling interpretation. Positive $\kappa_x$ and $\kappa_z$ imply mean reversion in the latent-variance and realized-measure states. A positive $\phi$ associates higher latent variance with a higher realized-measure level, while a positive $\gamma$ implies that increases in $z_t$ raise the drift of $x_t$, consistent with the measurement-driven role of realized measures~\citep{Hansen2012}. The condition $\gamma\phi<\kappa_x$ therefore permits positive two-way feedback while limiting its effective strength so that it remains below the latent-variance mean-reversion force needed for stability.

\subsection{Continuous-Time Dependence Structure}
\label{sec:ct_dependence}
The model contains Brownian innovations for returns, latent log-variance, the realized-measure state, and measurement noise. We assume that low return innovations are contemporaneously correlated with innovations in both volatility-related states:
\begin{equation}
    d\langle W^Y,W^x\rangle_t
    =
    \rho_{Y\!x}\,dt,
    \qquad
    d\langle W^Y,W^z\rangle_t
    =
    \rho_{Y\!z}\,dt.
    \label{eq:return_state_dependence}
\end{equation}
The parameters $\rho_{Y\!x}$ and $\rho_{Y\!z}$ jointly provide a parsimonious first-order representation of return-linked dependence. When these correlations are negative, negative return shocks are associated with positive innovations in volatility-related states. In contrast, positive return shocks tend to induce downward innovations relative to their baseline dynamics. Rather than introducing a separate discrete-time asymmetry function \citep{Hansen2012}, the proposed framework represents these effects through the joint Brownian dependence structure. We assume $d\langle W^x,W^z\rangle_t = 0$ to isolate the return-linked dependence channels. The two states nevertheless remain dynamically coupled through their drift interactions and feedback mechanism. In addition, we set $d\langle W^Y,U\rangle_t
=
d\langle W^x,U\rangle_t
=
d\langle W^z,U\rangle_t
=
0$, so that $U_t$ represents measurement-specific noise independent of other innovations~\citep{Hansen2012,BekiermanGribisch2016}.

Thus, for $d\mathbf{W}_t
    =
    \begin{pmatrix}
        dW_t^Y &
        dW_t^x &
        dW_t^z &
        dU_t
    \end{pmatrix}^{\top}$
the instantaneous quadratic-covariation matrix satisfies
\begin{equation}
    \mathbb{E}
    \left[
        d\mathbf{W}_t d\mathbf{W}_t^{\top}
    \right]
    =
    \mathbf{C}\,dt,
    \qquad
    \mathbf{C}
    =
    \begin{pmatrix}
        1         & \rho_{Y\!x} & \rho_{Y\!z} & 0 \\
        \rho_{Y\!x} & 1         & 0         & 0 \\
        \rho_{Y\!z} & 0         & 1         & 0 \\
        0         & 0         & 0         & 1
    \end{pmatrix}.
    \label{eq:brownian_corr}
\end{equation}

The parameters are constrained so that $\mathbf{C}$ is positive semidefinite, which in this specification requires
\begin{equation}
    \rho_{Y\!x}^2+\rho_{Y\!z}^2\leq 1.
\end{equation}
Implementation details are provided in Appendix~\ref{app:correlated_brownian}, and the predictive contribution of the joint return-linked dependence channel is examined in Section~\ref{ablationb}.

\subsection{Simulation-Based Predictive Learning}
\label{sec:predictive_learning}
The proposed SDE does not generally admit a tractable one-step predictive density for the joint daily return and realized-measure outcome. We therefore approximate this predictive distribution using Monte Carlo simulation\citep{KloedenPlaten1992, Glasserman2004}. For each training pair $(n,n+1)$, the Monte Carlo samples are
\begin{equation}
    q_{n+1}^{(m)}
    =
    \begin{pmatrix}
        r_{n+1}^{(m)}\\
        \log R_{n+1}^{(m)}
    \end{pmatrix},
    \qquad m=1,\ldots,M.
\end{equation}
At observation time $n+1$, the simulated return and realized-measure state are compared with the observed return and log-RV in $q_{n+1}^{\mathrm{obs}}$, while
instantaneous variance $s_t$ remains latent.

The simulated samples are used to estimate the predictive mean and covariance,
\begin{align}
    \widehat{\boldsymbol{\mu}}_{n+1}
    &=
    \frac{1}{M}\sum_{m=1}^{M}q_{n+1}^{(m)},\\
    \widehat{\boldsymbol{\Sigma}}_{n+1}
    &=
    \frac{1}{M-1}
    \sum_{m=1}^{M}
    \left(q_{n+1}^{(m)}-\widehat{\boldsymbol{\mu}}_{n+1}\right)
    \left(q_{n+1}^{(m)}-\widehat{\boldsymbol{\mu}}_{n+1}\right)^{\top}.
\end{align}
The corresponding observed outcome is 
\begin{equation}
    q_{n+1}^{\mathrm{obs}}
    =
    \begin{pmatrix}
        r_{n+1}\\
        \log R_{n+1}
    \end{pmatrix}.
\end{equation}

Model parameters are estimated by minimizing the Gaussian quasi-negative log-likelihood \citep{BollerslevWooldridge1992}: 
\begin{equation}
    \mathcal{L}_{n}
    =
    \frac{1}{2}
    \left[
        2\log(2\pi)
        +\log\left|\widehat{\boldsymbol{\Sigma}}_{n+1}\right|
        +
        \left(q_{n+1}^{\mathrm{obs}}
        -\widehat{\boldsymbol{\mu}}_{n+1}\right)^{\top}
        \widehat{\boldsymbol{\Sigma}}_{n+1}^{-1}
        \left(q_{n+1}^{\mathrm{obs}}
        -\widehat{\boldsymbol{\mu}}_{n+1}\right)
    \right].
    \label{eq:joint_qnll}
\end{equation}
This criterion evaluates the joint one-step-ahead predictive mean and covariance of returns and realized measures using a Gaussian quasi-score. It is a simulation-based predictive criterion rather than the exact observed-data likelihood of the latent continuous-time system.

The objective is averaged over each mini-batch and optimized by differentiating through the simulated SDE paths. Numerical details are provided in Appendix~\ref{Appendix:experimental details}.

\subsection{Simulation-Based Integrated-Variance Forecasting}
\label{sec:mc_inference}
Our forecast target is the one-day integrated variance $H_{n+1}$ in Eq~\eqref{eq:integrated_variance}, rather than the terminal instantaneous variance $s_{n+1}$, because $H_{n+1}$ corresponds directly to the daily forecasting horizon.

For a continuous log-price diffusion, integrated variance equals the quadratic variation of returns over the day. Under standard regularity conditions, realized variance converges in probability to quadratic variation as the intraday sampling frequency increases~\citep{Andersen2001, BarndorffNielsen2002}. Hence, $R_{n+1}$ provides an observable ex post proxy for $H_{n+1}$.

At each forecast origin $n$, conditional on states initialized using information available through time $n$, we simulate $M$ paths of the fitted SDE over $[n,n+1]$. These Monte Carlo paths approximate the predictive distribution implied by the fitted model. For each simulated path, the one-day integrated variance is
\begin{equation}
    H_{n+1}^{(m)}
    =
    \int_n^{n+1}
    \exp\!\left(x_t^{(m)}\right)\,dt.
    \label{eq:mc_integrated_variance}
\end{equation}
where $\exp(x_t^{(m)})$ is the simulated instantaneous variance along path $m$. The point forecast of next-day integrated variance is then obtained by averaging across the Monte Carlo samples,
\begin{equation}
    \widehat{H}_{n+1|n}
    =
    \frac{1}{M}
    \sum_{m=1}^{M}
    H_{n+1}^{(m)}.
    \label{eq:mc_variance_forecast}
\end{equation}

State initialization and numerical simulation details are reported in Appendix~\ref{Appendix:experimental details}.

\section{Experiments}
\subsection{Experimental Design}
\label{sec:experimental_design}
We evaluate the proposed model across 17 international equity indices using a 21-trading-day rolling-window design, approximately corresponding to monthly re-estimation. The initial estimation window contains approximately the first 50\% of each index. At each re-estimation date, the model is updated using the current rolling window and then held fixed over the following period to generate daily one-day-ahead integrated-variance forecasts using only information available at each forecast origin. The window is subsequently advanced, and the procedure is repeated. The implementation is available at \url{https://github.com/Haley-xixi/realized-garch-sde-code}.

The proposed SDE is compared with GARCH, GARCH-It\^{o}, Realized GARCH, and HEAVY. Forecasts are evaluated out of sample using QLIKE as the primary variance-loss criterion because it is robust to noise in realized-volatility proxies \citep{HansenLunde2005, Patton2011}. MSE${\mathrm{var}}$ and MAE${\mathrm{var}}$ are reported as additional metrics to measure forecast errors on the variance scale. Statistical significance of predictive differences is assessed using Diebold-Mariano (DM) tests \citep{DieboldMariano1995} and the Model Confidence Set (MCS) procedure \citep{Hansen2011}. 

In addition, we conduct controlled ablations to identify the contributions of realized-measure feedback and the correlated Brownian dependence structure. We also examine sensitivity to Monte Carlo sample size, SDE discretization, and random initialization. 
Additional implementation and evaluation details are provided in Appendix~\ref{Appendix:experimental details}.

\subsection{Forecasting Performance and Statistical Comparison}
\label{sec:forecast_comparison}
\begin{figure*}[htbp]
    \centering
    \begin{subfigure}[t]{0.32\textwidth}
        \centering
        \includegraphics[width=\linewidth]{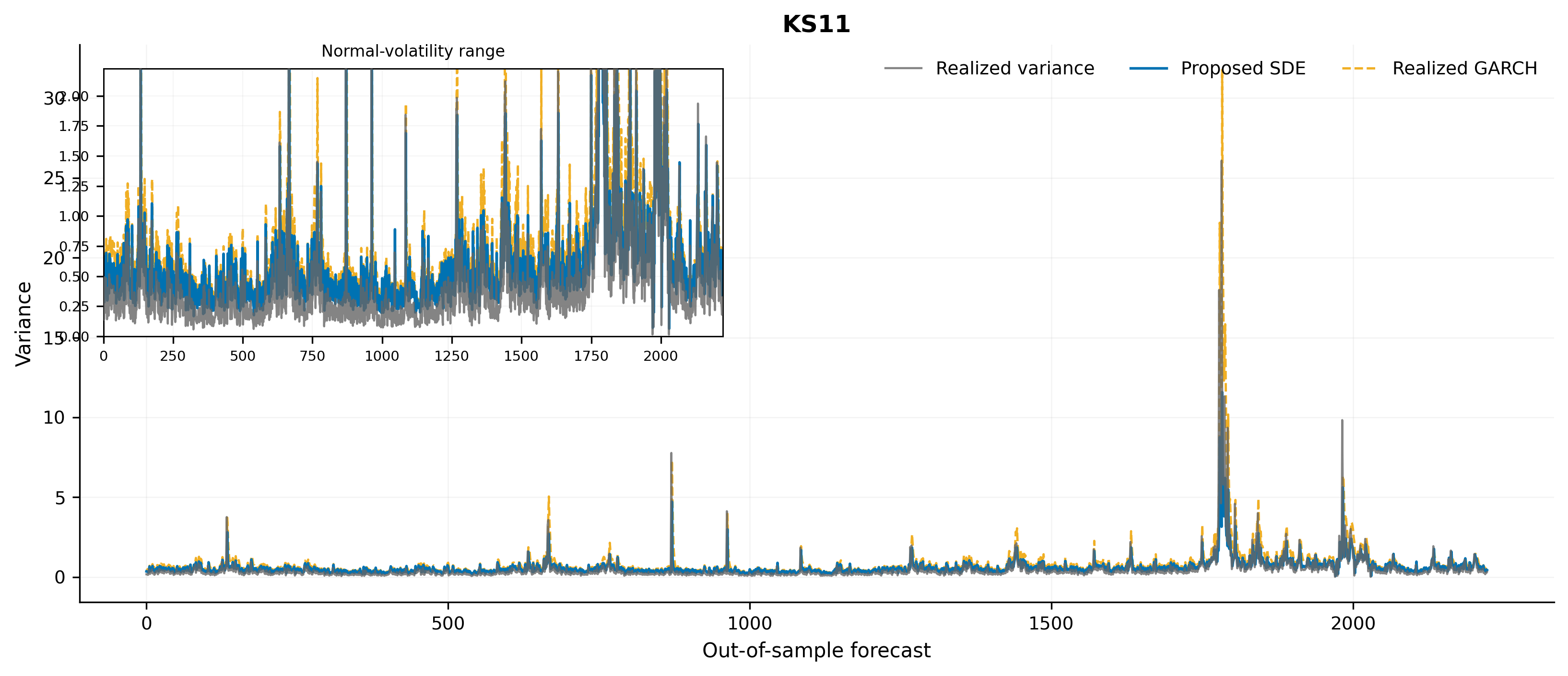}
        \caption{KS11}
        \label{fig:oos_forecasts_ks11}
    \end{subfigure}
    \hfill
    \begin{subfigure}[t]{0.32\textwidth}
        \centering
        \includegraphics[width=\linewidth]{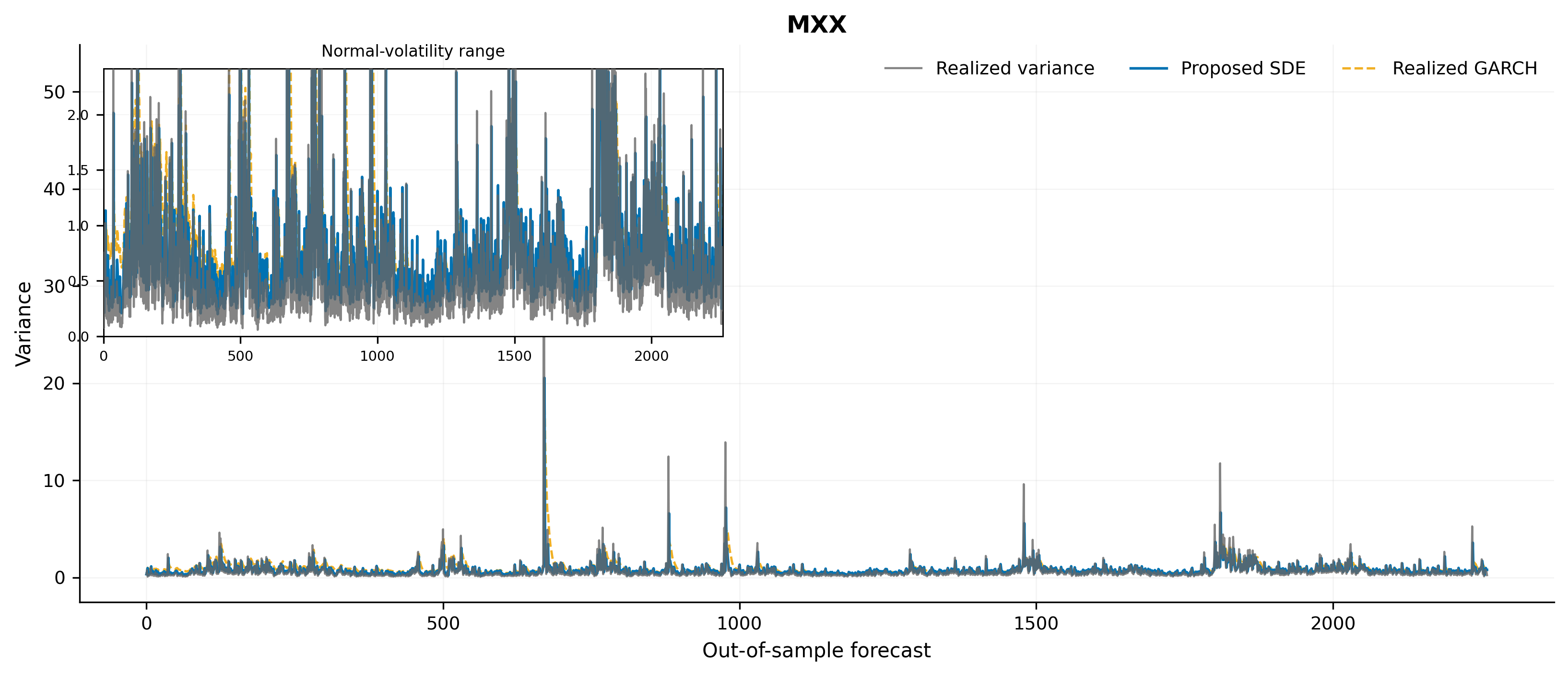}
        \caption{MXX}
        \label{fig:oos_forecasts_mxx}
    \end{subfigure}
    \hfill
    \begin{subfigure}[t]{0.32\textwidth}
        \centering
        \includegraphics[width=\linewidth]{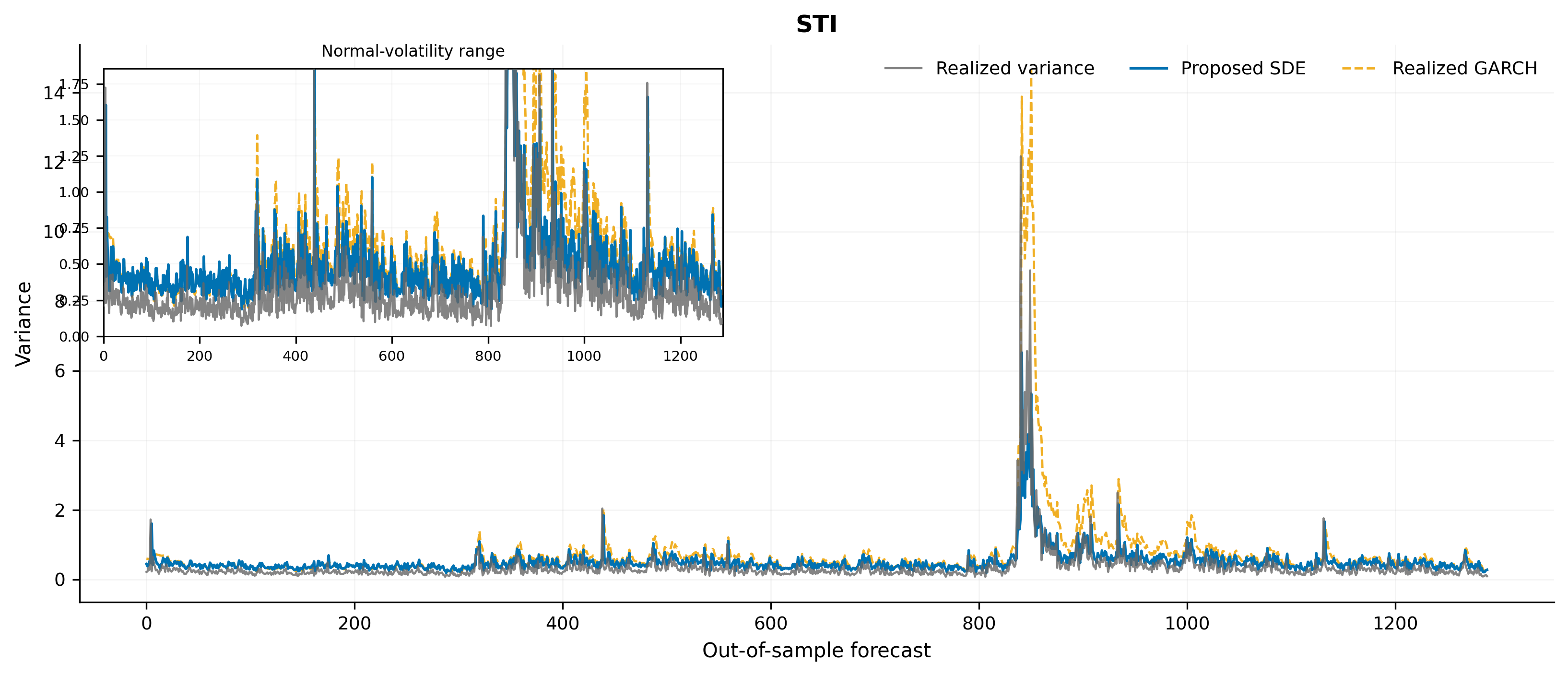}
        \caption{STI}
        \label{fig:oos_forecasts_sti}
    \end{subfigure}

    \caption{One-step-ahead variance forecasts of the proposed SDE and
    Realized GARCH for representative equity indices.}
    \label{fig:sde_realgarch_forecasts}
\end{figure*}

Figure~\ref{fig:sde_realgarch_forecasts} illustrates forecast behavior for KS11, MXX, and STI. Compared with Realized GARCH, the proposed SDE generally tracks RV more closely and exhibits less overshooting during several high-volatility episodes.  

To compare multiple models jointly, we use the MCS procedure
of~\citet{Hansen2011} with
\[
\mathcal{M}_0
=
\{
\text{proposed SDE},
\text{GARCH-It\^{o}},
\text{GARCH},
\text{Realized-GARCH},
\text{HEAVY}
\}.
\]

\begin{table*}[t]
\centering
\caption{
Out-of-sample integrated-variance forecasting performance evaluated against
raw realized variance across 17 equity indices.
Models include the proposed SDE (Prop.), GARCH, GARCH-It\^{o} (G-It\^{o}),
Realized GARCH (RGARCH), HEAVY, and HAR-QML (HAR).
Lower values indicate better performance.
Bold and underlined values denote the best and second-best models,
respectively. Shading indicates inclusion in the 90\% MCS.
}
\label{tab:oos_forecast_performance}
\scriptsize
\setlength{\tabcolsep}{2.2pt}
\renewcommand{\arraystretch}{0.95}

\resizebox{\textwidth}{!}{
\begin{tabular}{lrrrrrrrrrrrrrrrrrr}
\toprule
& \multicolumn{6}{c}{QLIKE}
& \multicolumn{6}{c}{MSE$_{\mathrm{var}}$}
& \multicolumn{6}{c}{MAE$_{\mathrm{var}}$} \\
\cmidrule(lr){2-7}
\cmidrule(lr){8-13}
\cmidrule(lr){14-19}

Index
& Prop. & GARCH & G-It\^{o} & RGARCH & HEAVY & HAR
& Prop. & GARCH & G-It\^{o} & RGARCH & HEAVY & HAR
& Prop. & GARCH & G-It\^{o} & RGARCH & HEAVY & HAR \\
\midrule

SPX
& \cellcolor{gray!20}\textbf{0.0709}
& 0.2015
& 0.1796
& 0.1131
& \cellcolor{gray!20}\underline{0.0743}
& 0.0783
& \cellcolor{gray!20}\underline{2.6538}
& \cellcolor{gray!20}4.3217
& \cellcolor{gray!20}4.1299
& \cellcolor{gray!20}\textbf{2.2426}
& \cellcolor{gray!20}3.1888
& \cellcolor{gray!20}3.5372
& \cellcolor{gray!20}\textbf{0.4285}
& 0.6539
& 0.6358
& \underline{0.4740}
& 0.5256
& 0.5393 \\

SSMI
& \cellcolor{gray!20}\textbf{0.1724}
& 0.2827
& 0.2886
& 0.2987
& \underline{0.2032}
& 0.2061
& \cellcolor{gray!20}\textbf{3.5165}
& \cellcolor{gray!20}\underline{3.7980}
& \cellcolor{gray!20}3.9539
& \cellcolor{gray!20}4.0661
& \cellcolor{gray!20}7.9464
& \cellcolor{gray!20}8.5839
& \cellcolor{gray!20}\textbf{0.3812}
& 0.6035
& 0.6175
& \underline{0.5408}
& 0.5994
& 0.6047 \\

STI
& \cellcolor{gray!20}\textbf{-0.1208}
& 0.0072
& 0.0813
& -0.0396
& -0.0414
& \underline{-0.0571}
& \cellcolor{gray!20}\textbf{0.2271}
& \cellcolor{gray!20}1.9968
& \cellcolor{gray!20}\underline{0.7090}
& \cellcolor{gray!20}0.8385
& \cellcolor{gray!20}1.1152
& \cellcolor{gray!20}1.4309
& \cellcolor{gray!20}\textbf{0.2367}
& 0.5607
& 0.4999
& \underline{0.4190}
& 0.4260
& 0.4493 \\

OMXSPI
& \cellcolor{gray!20}\textbf{0.0941}
& 0.3540
& \underline{0.2605}
& 0.3527
& 0.2687
& 0.2714
& \cellcolor{gray!20}\textbf{0.4805}
& \cellcolor{gray!20}1.8909
& \cellcolor{gray!20}\underline{1.1743}
& \cellcolor{gray!20}1.3234
& \cellcolor{gray!20}1.9251
& \cellcolor{gray!20}1.8188
& \cellcolor{gray!20}\textbf{0.3013}
& 0.7743
& \underline{0.5892}
& 0.6607
& 0.6434
& 0.6429 \\

KS11
& \cellcolor{gray!20}\textbf{0.0713}
& 0.1846
& 0.1888
& 0.1667
& 0.1614
& \underline{0.1581}
& \cellcolor{gray!20}\textbf{0.5928}
& \cellcolor{gray!20}1.2124
& \cellcolor{gray!20}\underline{0.9317}
& \cellcolor{gray!20}1.1464
& \cellcolor{gray!20}1.4271
& \cellcolor{gray!20}1.2803
& \cellcolor{gray!20}\textbf{0.2840}
& 0.4919
& \underline{0.4650}
& 0.4802
& 0.5165
& 0.5058 \\

OMXHPI
& \cellcolor{gray!20}\textbf{0.1767}
& 0.4232
& \underline{0.3108}
& 0.3757
& 0.3735
& 0.3753
& \cellcolor{gray!20}\textbf{0.3433}
& 1.4279
& 1.0842
& \underline{0.9082}
& 1.2894
& 1.2828
& \cellcolor{gray!20}\textbf{0.3106}
& 0.7591
& \underline{0.5786}
& 0.6321
& 0.6736
& 0.6754 \\

SSEC
& \cellcolor{gray!20}\textbf{0.7730}
& 0.9105
& \underline{0.8521}
& 0.8846
& 0.8669
& 0.8537
& \cellcolor{gray!20}\underline{4.4282}
& \cellcolor{gray!20}5.3415
& \cellcolor{gray!20}\textbf{4.3308}
& \cellcolor{gray!20}4.5293
& \cellcolor{gray!20}5.0989
& \cellcolor{gray!20}4.6997
& \cellcolor{gray!20}\textbf{0.6749}
& 1.1528
& \underline{1.0193}
& 1.0513
& 1.0598
& 1.0325 \\

KSE
& \cellcolor{gray!20}\textbf{0.3161}
& 0.4766
& 0.4703
& 0.4513
& 0.4764
& \underline{0.4435}
& \cellcolor{gray!20}\textbf{1.0611}
& \underline{1.3870}
& 1.4036
& 1.4956
& 1.8165
& 1.8841
& \cellcolor{gray!20}\textbf{0.3866}
& 0.7110
& 0.7029
& \underline{0.6587}
& 0.7213
& 0.6825 \\

MXX
& \cellcolor{gray!20}\textbf{0.4577}
& \underline{0.5006}
& 0.5696
& 0.5149
& 0.5104
& 0.5130
& \cellcolor{gray!20}\textbf{1.8498}
& \cellcolor{gray!20}2.1519
& \cellcolor{gray!20}2.6185
& \cellcolor{gray!20}2.1114
& \cellcolor{gray!20}\underline{2.0700}
& \cellcolor{gray!20}2.1902
& \cellcolor{gray!20}\textbf{0.4261}
& \underline{0.5612}
& 0.7160
& 0.5669
& 0.5686
& 0.5754 \\

IXIC
& \cellcolor{gray!20}\textbf{0.2686}
& 0.4876
& 0.4623
& 0.3988
& 0.3941
& \underline{0.3921}
& \cellcolor{gray!20}\textbf{2.2612}
& \cellcolor{gray!20}3.7000
& \cellcolor{gray!20}\underline{3.0887}
& \cellcolor{gray!20}4.1079
& \cellcolor{gray!20}6.6113
& \cellcolor{gray!20}6.8050
& \cellcolor{gray!20}\textbf{0.4341}
& 0.8332
& 0.7765
& \underline{0.7681}
& 0.8332
& 0.8370 \\

OMXC20
& \cellcolor{gray!20}\textbf{0.7772}
& 0.8602
& \underline{0.8385}
& 0.8596
& 0.8405
& 0.8400
& \cellcolor{gray!20}5.1722
& \cellcolor{gray!20}\underline{4.9324}
& \cellcolor{gray!20}\textbf{4.8671}
& \cellcolor{gray!20}5.5860
& 7.1864
& 7.1971
& \cellcolor{gray!20}\textbf{0.5813}
& 0.7856
& \underline{0.7157}
& 0.8299
& 0.8825
& 0.8825 \\

STOXX50E
& 0.8867
& 0.8201
& 0.8101
& 0.8293
& \cellcolor{gray!20}\textbf{0.7462}
& \cellcolor{gray!20}\underline{0.7474}
& \cellcolor{gray!20}\textbf{4.0608}
& \cellcolor{gray!20}4.4580
& \cellcolor{gray!20}\underline{4.1166}
& \cellcolor{gray!20}4.5844
& \cellcolor{gray!20}4.6357
& \cellcolor{gray!20}4.5044
& \cellcolor{gray!20}\textbf{0.5899}
& 0.8743
& 0.8413
& 0.7823
& 0.7497
& \underline{0.7450} \\

SMSI
& \cellcolor{gray!20}\textbf{0.7595}
& 0.8975
& 0.9158
& 0.9082
& 0.8335
& \underline{0.8301}
& \cellcolor{gray!20}\textbf{4.6071}
& \cellcolor{gray!20}\underline{6.1262}
& \cellcolor{gray!20}7.0285
& \cellcolor{gray!20}6.2267
& \cellcolor{gray!20}6.4199
& \cellcolor{gray!20}6.3419
& \cellcolor{gray!20}\textbf{0.5146}
& 0.9906
& 1.0668
& 0.9218
& 0.8655
& \underline{0.8523} \\

N225
& \cellcolor{gray!20}\textbf{0.4498}
& 0.7945
& 0.7594
& 0.9133
& 0.7402
& \underline{0.7400}
& \cellcolor{gray!20}\textbf{1.9749}
& 4.1690
& 4.1436
& \underline{3.3694}
& 6.8496
& 6.6623
& \cellcolor{gray!20}\textbf{0.4754}
& 1.3310
& \underline{1.2802}
& 1.3138
& 1.3603
& 1.3466 \\

OSEAX
& \cellcolor{gray!20}\textbf{0.6341}
& 0.6732
& 0.6971
& 0.7061
& \cellcolor{gray!20}0.6470
& \cellcolor{gray!20}\underline{0.6453}
& \cellcolor{gray!20}20.6389
& \cellcolor{gray!20}\underline{19.1486}
& \cellcolor{gray!20}\textbf{18.8334}
& \cellcolor{gray!20}19.8459
& \cellcolor{gray!20}21.6862
& \cellcolor{gray!20}21.8278
& \cellcolor{gray!20}\textbf{0.6978}
& \underline{0.8189}
& 0.8912
& 0.9211
& 0.9289
& 0.9331 \\

RUT
& \cellcolor{gray!20}\textbf{0.3307}
& 0.6484
& 0.7082
& 0.6388
& 0.6290
& \underline{0.6259}
& \cellcolor{gray!20}\textbf{1.3142}
& 7.0445
& 10.6781
& \underline{2.8758}
& 5.4931
& 5.4987
& \cellcolor{gray!20}\textbf{0.4272}
& 1.1307
& 1.3200
& \underline{1.0188}
& 1.1467
& 1.1426 \\

NSEI
& 0.4990
& 0.4668
& \cellcolor{gray!20}0.4297
& 0.4495
& \cellcolor{gray!20}\underline{0.4083}
& \cellcolor{gray!20}\textbf{0.4067}
& \cellcolor{gray!20}\underline{5.0865}
& \cellcolor{gray!20}6.1535
& \cellcolor{gray!20}\textbf{4.5578}
& \cellcolor{gray!20}5.5442
& \cellcolor{gray!20}6.0896
& \cellcolor{gray!20}6.8380
& \cellcolor{gray!20}\textbf{0.4379}
& 0.7592
& 0.6461
& 0.6596
& \underline{0.6377}
& 0.6466 \\

\midrule
Best count
& 15 & 0 & 0 & 0 & 1 & 1
& 12 & 0 & 4 & 1 & 0 & 0
& 17 & 0 & 0 & 0 & 0 & 0 \\

MCS count
& 15 & 0 & 1 & 0 & 4 & 3
& 17 & 13 & 13 & 13 & 12 & 12
& 17 & 0 & 0 & 0 & 0 & 0 \\

\bottomrule
\end{tabular}
}
\end{table*}

Table~\ref{tab:oos_forecast_performance} reports the surviving 90\% MCS across the competing models, where the included models are not significantly inferior to the superior set at the level of 10\% \citep{Hansen2011}. The results provide strong evidence that the proposed SDE is consistently competitive for integrated variance forecasting. It achieves the lowest loss for 15 of 17 indices under QLIKE, 12 under MSE${\mathrm{var}}$, and all 17 indices under MAE${\mathrm{var}}$. It is also retained in the 90\% MCS for 15 indices under QLIKE and all 17 under both MSE${\mathrm{var}}$ and MAE${\mathrm{var}}$. Together, the best-count and MCS results indicate that the proposed SDE performs competitively across a wide range of equity markets and evaluation criteria, and this performance is not confined to particular markets or loss functions.

\subsection{Ablation Study}
\label{sec:ablation_design}
To address component-level contribution questions, we design two controlled ablation variants:
(i) removing realized-measure feedback by setting $\gamma=0$; and
(ii) removing return-linked stochastic dependence by imposing
$\rho_{Y\!x}=\rho_{Y\!z}=0$.
In each ablation, all unrestricted parameters are re-estimated while the remaining experimental settings are held fixed. We then conduct pairwise DM tests between the full model and each ablation under QLIKE, MSE${\mathrm{var}}$, and MAE${\mathrm{var}}$. These comparisons assess whether realized-measure feedback and first-order contemporaneous return-volatility dependence contribute additional forecasting value. Complete index-level DM test results for both ablations are reported in Appendix~\ref{app:ablation_results}.

\paragraph{A. Does realized-measure feedback improve forecasting?}
\label{ablationa}

\begin{wraptable}{r}{0.58\textwidth}
\centering 
\caption{Realized-Measure Feedback Ablation: Pairwise DM Results}
\label{tab:dm_gamma_ablation}
\tiny
\begin{tabular}{lcccc}
\toprule
& \multicolumn{2}{c}{Lower Mean Loss}
& \multicolumn{2}{c}{Significantly Better (10\%)} \\
\cmidrule(lr){2-3}
\cmidrule(lr){4-5}
Metric
& Full SDE
& $\gamma=0$
& Full SDE
& $\gamma=0$ \\
\midrule

QLIKE
& \textbf{14/17}
& 3/17
& \textbf{11/17}
& 0/17 \\

MSE$_{\mathrm{var}}$
& \textbf{13/17}
& 4/17
& \textbf{2/17}
& 1/17 \\

MAE$_{\mathrm{var}}$
& \textbf{17/17}
& 0/17
& \textbf{13/17}
& 0/17 \\

\bottomrule
\end{tabular}

\vspace{0.4em}
\begin{minipage}{0.95\linewidth}
\footnotesize
\textit{Notes:}
Entries are counts across 17 equity indices. ``Lower Mean Loss'' reports the number of indices for which each specification achieves the lower average loss. ``Significantly Better (10\%)'' reports the number of indices for which each specification has significantly lower loss than its alternative according to the DM test at the 10\% level. Bold entries indicate the larger count within each pairwise comparison.
\end{minipage}
\end{wraptable}

We compare the Full SDE, in which $\gamma$ is estimated, with a restricted specification that imposes $\gamma=0$. This restriction removes the direct realized-measure contribution to the latent log-variance drift while leaving the remaining state dynamics and dependence structure unchanged.

According to Table~\ref{tab:dm_gamma_ablation}, the Full SDE achieves lower QLIKE and MAE${\mathrm{var}}$ in 14 and 17 of 17 indices, respectively, with significant improvements in 11 and 13 markets. For MSE${\mathrm{var}}$, it achieves a lower loss in 13 markets, although only two differences are significant. This pattern suggests that allowing observed RV information to feed directly into latent volatility dynamics helps the model track evolution of latent variance, providing incremental information for integrated-variance forecasting. We assess return-linked stochastic dependence by comparing Full SDE with an Independent specification.

\paragraph{B. Does return-linked stochastic dependence improve predictive performance?}
\label{ablationb}
We assess return-linked stochastic dependence by comparing the Full SDE with an Independent specification imposing $ \rho_{Y\!x} = \rho_{Y\!z}  =  0$, while keeping all other model components unchanged. This ablation evaluates whether allowing contemporaneous return innovations to be jointly correlated with innovations in the latent-volatility and realized-measure states contributes to predictive performance.

\begin{wraptable}{r}{0.65\textwidth}
\centering
\caption{Return-linked Stochastic Dependence Ablation: Pairwise DM Results}
\label{tab:dm_brownian_dependence}
\tiny
\begin{tabular}{lcccc}
\toprule
&
\multicolumn{2}{c}{Lower Mean Loss}
&
\multicolumn{2}{c}{Significantly Better (10\%)}
\\
\cmidrule(lr){2-3}
\cmidrule(lr){4-5}

Metric
& Full SDE
& Independent
& Full SDE
& Independent
\\
\midrule

QLIKE
& \textbf{10/17}
& 7/17
& \textbf{3/17}
& 0/17
\\

MSE$_{\mathrm{var}}$
& \textbf{12/17}
& 5/17
& 0/17
& 0/17
\\

MAE$_{\mathrm{var}}$
& \textbf{11/17}
& 6/17
& \textbf{1/17}
& 0/17
\\

\bottomrule
\end{tabular}

\vspace{0.3em}
\begin{minipage}{0.95\linewidth}
\footnotesize
\textit{Notes:}
Definitions follow Table~\ref{tab:dm_gamma_ablation}. Entries are counts across 17 equity indices, and bold entries indicate the larger count within each comparison.
\end{minipage}
\end{wraptable}

Importantly, the correlated Brownian structure captures only a parsimonious first-order contemporaneous dependence channel, rather than the full nonlinear return-measurement relationship. For example, Realized GARCH allows both
linear and quadratic effects of standardized return innovations through $\tau_1\varepsilon_n+\tau_2(\varepsilon_n^2-1)$ \citep{Hansen2012}. Our results therefore suggest that the linear instantaneous dependence channel provides some incremental forecasting value in Table~\ref{tab:dm_brownian_dependence}, but its contribution is weaker than that of realized-measure feedback.

\subsection{Sensitivity Analysis}
\label{sec:sensitivity}
We assess the robustness of the proposed model to key numerical and estimation choices, including the number of Monte Carlo samples, SDE discretization resolution, and random initialization \citep{KloedenPlaten1992,Glasserman2004}. Forecasting performance remains broadly stable across the alternative settings, indicating that the main empirical results are not driven by numerical or optimization choices. Detailed results are reported in Appendix~\ref{app:sensitivity}.

\section{Discussion}

\subsection{Main Findings}
Based on our empirical experiments, several conclusions emerge about the economic mechanisms underlying the proposed model. First, the results support using of realized-measure information as a direct input to continuous-time latent variance dynamics. RV aggregates intraday price variation and therefore provides information about volatility that is not fully contained in daily returns. By allowing the realized-measure state to enter directly into the drift of latent variance, the proposed model provides a mechanism through which realized-measure information can influence subsequent latent variance dynamics. The strong out-of-sample performance across markets suggests that this measurement-driven information remains useful for one-day-ahead integrated-variance forecasting when embedded within a continuous-time stochastic volatility system.

The realized-measure feedback ablation provides more direct evidence for this mechanism. Setting $\gamma=0$ asks whether the realized-measure state contains incremental information for future latent variance dynamics once the remaining stochastic volatility dynamics are retained. The  substantial deterioration under QLIKE and MAE$_{\mathrm{var}}$ indicates that the realized-measure state is not merely a contemporaneous proxy for latent variance. It also contains useful information for the subsequent evolution of the latent variance process. This provides a continuous-time interpretation of the measurement-driven intuition underlying Realized GARCH and HEAVY \citep{Hansen2012,Shephard2010}.

The return-linked dependence ablation tests whether contemporaneous return shocks provide additional information about innovations in volatility-related states. Correlated Brownian dependence yields modest gains with limited statistical significance, suggesting a complementary role relative to realized-measure feedback. This may reflect the parsimonious dependence structure: Brownian correlation captures only linear, sign-sensitive contemporaneous dependence, where negative and positive return shocks induce opposite-direction volatility innovations. However, it cannot capture nonlinear magnitude effects in which large shocks of either sign generate stronger responses. Thus,  the results indicate limited incremental value from this first-order dependence channel, rather than from general return-volatility asymmetry.  

\subsection{Limitations and Future Work}
\label{sec:limitations}
Several limitations qualify the empirical findings. First, the current study operates on a daily observation grid for comparability with the benchmark models, with high-frequency information entering only through daily realized variance. Consequently, the intraday evolution of the continuous-time states remains latent and is approximated through Monte Carlo simulation and Euler-Maruyama discretization \citep{KloedenPlaten1992,Glasserman2004}. Although the sensitivity analysis suggests that the main results are broadly stable, this introduces simulation and discretization error. Future work could incorporate intraday observations directly to update the state processes at a finer frequency and reduce reliance on simulated intraday paths.

Second, the current specification is deliberately parsimonious. The realized-measure feedback is linear, $g(z_t)=z_t$, and the dependence structure focuses on return-linked correlations while imposing
$d\langle W^x,W^z\rangle_t=0$. Richer nonlinear or learned feedback functions and more flexible dependence structures are natural extensions.

Finally, the empirical analysis focuses on one-day-ahead integrated-variance forecasting for equity indices. Whether the same mechanisms remain effective at longer horizons, for individual equities, or across other asset classes remains an open question. Extending the framework to broader forecasting settings would provide a useful test of its generality.

\bibliographystyle{unsrtnat}
\bibliography{references}

\clearpage
\appendix
\section{Related Work}\label{Appendix:A}

\subsection{Discrete-Time Volatility Models}

\paragraph{Return-based volatility models.}
A foundational approach to volatility modeling is to infer time-varying conditional variance from observed asset returns. The ARCH model of~\citet{Engle1982} captures volatility clustering by allowing current conditional variance to depend on past squared return innovations. GARCH extends this framework by additionally incorporating lagged conditional variance, providing a parsimonious representation of the persistence commonly observed in financial volatility~\citep{Bollerslev1986}. These models established the importance of modeling volatility as a dynamic conditional variance process and remain widely used benchmarks for volatility forecasting. However, because they rely primarily on daily return information and lagged conditional variances, they do not directly exploit the richer information on intraday price variation contained in high-frequency data~\citep {Andersen2001}.

\paragraph{Realized-measure-based volatility models.}
The increasing availability of high-frequency price data has enabled more direct measurement of volatility through realized measures. In a continuous diffusion setting, RV constructed from intraday returns provides a consistent estimate of daily integrated variance under standard conditions~\citep{Andersen2001, BarndorffNielsen2002}. This development has motivated a range of discrete-time models that explicitly exploit realized measures for volatility forecasting.

One prominent approach is the heterogeneous autoregressive model of RV (HAR-RV), which captures volatility persistence across daily, weekly, and monthly horizons through a simple autoregressive structure~\citep{Corsi2009}. The  HEAVY framework instead incorporates realized measures directly into the dynamics of conditional return variance, allowing high-frequency information to improve the responsiveness of volatility forecasts to new market information~\citep{Shephard2010}. Realized GARCH further develops this idea by jointly modeling returns, conditional variance, and realized measures through a measurement equation linking the observed realized measure to the conditional variance process~\citep{Hansen2012}. This augments the standard GARCH recursion with high-frequency intraday information while retaining its familiar dynamic structure.

These models establish the empirical value of realized measures for volatility modeling by supplementing daily returns with high-frequency information. Their dynamics, however, remain defined in discrete time. This motivates considering how the same realized-measure information can be represented in a continuous-time latent-variance framework.

\subsection{Continuous-Time Volatility Models}

Continuous-time volatility models provide an alternative to discrete-time specifications by representing asset prices and latent volatility through SDEs. In stochastic volatility models, latent variance evolves as a stochastic process, allowing its dynamics to be specified directly in continuous time. A canonical example is the Heston model, which specifies variance as a mean-reverting diffusion and allows its innovations to be correlated with asset-price innovations, thereby capturing the leverage dependence between returns and volatility~\citep{Heston1993}.

A related literature examines the continuous-time limits of discrete-time GARCH models.~\citet{Nelson1990} showed that, when the observation interval shrinks and the GARCH parameters are scaled appropriately, the discrete conditional variance recursion converges to a diffusion process. This result provides a formal diffusion approximation to discrete-time GARCH dynamics.~\citet{Duan1997} extended this approach to a wider class of GARCH specifications, including models with more general asymmetric and nonlinear variance dynamics, and derived the corresponding diffusion limits.

More recently, the unified GARCH-It\^o framework embeds GARCH-type volatility dynamics in a continuous-time Itô process, allowing volatility to evolve continuously within each trading day while retaining a GARCH-type structure at the daily frequency~\citep{KimWang2016}. The realized GARCH-Itô model extends this framework by embedding realized-GARCH-type dynamics into a continuous-time jump-diffusion, allowing high-frequency measures of integrated and jump variation to enter the daily volatility dynamics~\citep{Song2021}. Together, these models provide a link between continuous-time price dynamics and discrete-time volatility updating.

These approaches demonstrate the value of modeling volatility as a continuous-time stochastic process. However, our specification differs by introducing a jointly evolving continuous-time realized-measure state that feeds directly into the drift of the latent variance process.

\section{Structural Motivation from Realized GARCH}
\label{Appendix:B}
This appendix explains how the discrete-time Realized GARCH framework~\citep{Hansen2012} informs the design of the proposed structured SDE.

Our model is not intended as a continuous-time extension or diffusion limit of Realized GARCH. Rather, we use its measurement-driven structure as guidance for constructing the continuous-time system. In particular, we do not preserve the individual nonlinear innovation terms in the Realized GARCH measurement equation. Instead, we represent their stochastic contribution more parsimoniously through diffusion in the realized-measure state. The resulting specification retains the main economic relationships suggested by Realized GARCH while allowing both latent variance and realized-measure information to evolve jointly in continuous time.

\subsection{Structural Motivation from Realized GARCH}

Consider the standard log-linear Realized GARCH structure
\begin{align}
    r_n
    &=
    \sqrt{s_n}\,\varepsilon_n,
    \label{eq:app_rg_return}
    \\
    x_n
    &=
    \omega
    +
    \beta x_{n-1}
    +
    \gamma z_{n-1},
    \label{eq:app_rg_variance}
    \\
    z_n
    &=
    \xi
    +
    \phi x_n
    +
    \tau_1\varepsilon_n
    +
    \tau_2(\varepsilon_n^2-1)
    +
    u_n,
    \label{eq:app_rg_measurement}
\end{align}
where $x_n=\log s_n$ and $z_n=\log R_n$, with $s_n$ denoting the
latent conditional variance and $R_n$ the observed realized measure on
trading day $n$~\citep{Hansen2012}. We use these relations as structural
motivation for the corresponding continuous-time components developed below.

\subsection{Return Dynamics}

Equation~\eqref{eq:app_rg_return} represents the return innovation over a discrete trading-day interval. To formulate an analogous local continuous-time representation, we interpret $s_t$ as the instantaneous variance rate~\citep{Heston1993}. The variance accumulated over $[t,t+\Delta t]$ is
\begin{align}
    \int_t^{t+\Delta t}s_u\,du.
\end{align}
For sufficiently small $\Delta t$, $s_u\approx s_t$ over the interval, so
\begin{align}
    \int_t^{t+\Delta t}s_u\,du
    \approx
    \int_t^{t+\Delta t}s_t\,du
    =
    s_t\Delta t.
\end{align}
The corresponding log-price increment can therefore be represented locally
as
\begin{align}
    Y_{t+\Delta t}-Y_t=
    \Delta Y_t
    \approx
    \sqrt{s_t\Delta t}\,\varepsilon,
    \qquad
    \varepsilon\sim\mathcal N(0,1),
    \label{eq:app_short_interval_return}
\end{align}
consistent with the usual local Gaussian approximation for It\^o
diffusions~\citep{Oksendal2003,KloedenPlaten1992}.

A Brownian increment over the same interval satisfies
\begin{align}
    \Delta W_t^Y
    =
    W_{t+\Delta t}^Y-W_t^Y
    \sim
    \mathcal{N}(0,\Delta t).
    \label{eq:brownian_increment_return}
\end{align}
Since
\begin{align}
    \sqrt{\Delta t}\,\varepsilon
    \sim
    \mathcal{N}(0,\Delta t),
\end{align}
we have
\begin{align}
    \sqrt{\Delta t}\,\varepsilon
    \overset{d}{=}
    \Delta W_t^Y,
\end{align}
where $\overset{d}{=}$ denotes equality in distribution. Hence, the
short-interval Gaussian innovation in
Eq.~\eqref{eq:app_short_interval_return} can be represented by a Brownian
increment,
\begin{align}
    \Delta Y_t
    \approx
    \sqrt{s_t}\,\Delta W_t^Y.
\end{align}
This local representation motivates the continuous-time log-price SDE~\citep{Heston1993}
\begin{align}
    dY_t
    =
    \sqrt{s_t}\,dW_t^Y.
    \label{eq:app_return_sde_general}
\end{align}

\subsection{Latent Volatility Dynamics}
The discrete-time log-variance recursion in
Eq.~\eqref{eq:app_rg_variance} can be rearranged as
\begin{align}
    x_n-x_{n-1}
    &=
    \omega
    -(1-\beta)x_{n-1}
    +\gamma z_{n-1}.
\end{align}
This formulation motivates two features of our continuous-time specification: mean reversion in latent log variance and direct feedback from the realized-measure state. Under the linear feedback specification $g(z) = z$, we adopt the following local Gaussian representation associated with It\^o diffusion for a short interval $[t,t+\Delta t]$ \citep{KloedenPlaten1992}. We therefore specify 
\begin{equation}
    \Delta x_t \approx \left[\kappa_x(\theta-x_t) +\gamma z_t\right]\Delta t+
    \eta_x\sqrt{\Delta t}\,\varepsilon_x,
    \qquad
    \varepsilon_x\sim\mathcal{N}(0,1),
    \label{eq:app_x_short_interval}
\end{equation}

where $\kappa_x$, $\theta$ and $\gamma$ are continuous-time model parameters. The notation reflects analogous structural roles rather than a direct parameter mapping from Realized GARCH.

Applying the same Brownian-scaling argument as for the return process,
\begin{align}
    \sqrt{\Delta t}\,\varepsilon_x
    \overset{d}{=}
    \Delta W_t^x,
\end{align}
so the short-interval change can be represented as
\begin{align}
    \Delta x_t
    \approx
    \left[
        \kappa_x(\theta-x_t)
        +
        \gamma z_t
    \right]\Delta t
    +
    \eta_x\Delta W_t^x.
\end{align}
This motivates the continuous-time specification
\begin{align}
    dx_t
    =
    \left[
        \kappa_x(\theta-x_t)
        +
        \gamma z_t
    \right]dt
    +
    \eta_x\,dW_t^x.
    \label{eq:app_x_sde_general}
\end{align}

To allow contemporaneous dependence between return and latent-variance innovations, we further specify
\begin{align}
    d\langle W^Y,W^x\rangle_t
    =
    \rho_{Y\!x}\,dt.
\end{align}

\subsection{Realized-Measure Dynamics}
The dynamics of the realized-measure state are motivated by the log-linear
measurement relation in Eq.~\eqref{eq:app_rg_measurement}. Under the standard Realized GARCH assumptions
\citep{Hansen2012}, the return innovation is standardized and the measurement error is mean-zero, so that
\begin{align}
    \mathbb{E}[\varepsilon_n\mid x_n]
    &=
    0,
    &
    \mathbb{E}[\varepsilon_n^2-1\mid x_n]
    &=
    0,
    &
    \mathbb{E}[u_n\mid x_n]
    &=
    0,
\end{align}
the innovation component
$\tau_1\varepsilon_n+\tau_2(\varepsilon_n^2-1)+u_n$
has zero conditional mean~\citep{Hansen2012}. Hence,
\begin{align}
    \mathbb{E}[z_n\mid x_n]
    =
    \xi+\phi x_n.
\end{align}

Motivated by this conditional relation, we take
\begin{equation}
    \mu_t = \xi+\phi x_t
\end{equation}
as a time-varying reference level for the continuous-time realized-measure state. We then introduce gradual adjustment toward this level by specifying
\begin{align}
    \mathbb{E}
    \left[
        z_{t+\Delta t}-z_t
        \mid \mathcal{F}_t
    \right]
    =
    \kappa_z(\mu_t-z_t)\Delta t
    +
    o(\Delta t).
    \label{eq:app_z_local_mean}
\end{align}
Dividing by $\Delta t$ and letting $\Delta t\to0$ gives
\begin{align}
    \lim_{\Delta t\to0}
    \frac{
        \mathbb{E}\!\left[
            z_{t+\Delta t}-z_t
            \mid \mathcal{F}_t
        \right]
    }{\Delta t}
    =
    \kappa_z(\mu_t-z_t),
\end{align}
which identifies the instantaneous drift of the realized-measure state.
Substituting $\mu_t=\xi+\phi x_t$ yields the mean reversion term
\begin{align}
    \kappa_z
    \left(
        \xi+\phi x_t-z_t
    \right).
\end{align}
It is positive when $z_t$ lies below $\xi+\phi x_t$ and vice versa, thereby pulling the realized-measure state toward the
time-varying level motivated by the Realized GARCH measurement relation.

The stochastic terms in the discrete measurement equation are not mapped
one-to-one into continuous-time processes. Instead, random variation in
$z_t$ is represented through diffusion components. We assume a low $W_t^z$ to be
correlated with the Brownian motion driving returns,
\begin{align}
    d\langle W^Y,W^z\rangle_t
    =
    \rho_{Y\!z}\,dt.
    \label{eq:app_Yz_covariation}
\end{align}
This provides a parsimonious continuous-time representation of the dependence between return innovations and realized-measure innovations.

Finally, $U_t$ is taken to be independent of the other Brownian motions, and
$\sigma_u\,dU_t$ captures additional measurement-specific variation in the
realized-measure state. The resulting dynamics are
\begin{align}
    dz_t
    =
    \kappa_z
    \left(
        \xi+\phi x_t-z_t
    \right)dt
    +
    \eta_z\,dW_t^z
    +
    \sigma_u\,dU_t.
    \label{eq:app_z_sde}
\end{align}

\subsection{State Initialization and Propagation}
\label{app:state_initialization}

For each one-step-ahead forecast formed at the end of trading day $n$,
the model state is initialized using only information available at the
corresponding forecast origin. The observed quantities available at
that time are the log-price $Y_n$ and the realized variance $R_n$.

Motivated by the role of realized variance as an observable
high-frequency proxy for latent integrated variance
\citep{andersen2003modeling}, we adopt a
proxy-based initialization for the latent variance state. Specifically,
the most recently observed realized variance is used to initialize both
the latent log-variance state and the realized-measure state:
\begin{equation}
    x_0^{(n)} = \log R_n,
    \qquad
    z_0^{(n)} = \log R_n.
\end{equation}
$R_n$ provides an observable starting value for the
otherwise latent variance state. For the realized-measure state,
$z_n=\log R_n$ follows directly from the specification
$R_t=\exp(z_t)$.

The forecast-origin state is therefore represented by the joint state
vector
\begin{equation}
    X_n
    =
    \left(
        Y_n,
        x_n,
        z_n
    \right).
\end{equation}
The three components are treated as a coupled state system. Conditional on $X_n$, the log-price state,
latent log-variance state, and realized-measure state evolve jointly
over $t\in(n,n+1]$ according to the coupled SDE dynamics in
Eqs.~\eqref{eq:new_1}--\eqref{eq:new_3}. Their joint propagation
allows contemporaneous dependence and feedback among the state
variables to operate throughout the one-day forecasting interval.

\section{Experimental Details}\label{Appendix:experimental details}
\subsection{Dataset}
\label{sec:dataset}

\paragraph{Data source and markets.}
We evaluate the proposed framework using daily returns and 5-minute RV measures for 17 major international equity indices from 2004 to 2022. We obtain the data from the Wharton Research Data Services (WRDS) database, which provides historical financial-market data and high-frequency realized measures used in empirical asset-pricing and volatility research. The sample covers equity markets across North America, Europe, and Asia-Pacific, providing a heterogeneous evaluation environment with substantial variation in market size, trading activity, and volatility dynamics.

The indices considered are the S\&P 500 (SPX), Swiss Market Index (SSMI), Straits Times Index (STI), OMX Stockholm All-Share Index (OMXSPI), Korea Composite Stock Price Index (KS11), OMX Helsinki All-Share Index (OMXHPI), Shanghai Composite Index (SSEC), Korea Stock Exchange Index (KSE), IPC Mexico Index (MXX), NASDAQ Composite Index (IXIC), OMX Copenhagen 20 Index (OMXC20), EURO STOXX 50 Index (STOXX50E), Madrid Stock Exchange General Index (SMSI), Nikkei 225 (N225), Oslo Stock Exchange All-Share Index (OSEAX), Russell 2000 (RUT), and NIFTY 50 (NSEI). Because trading calendars and data availability differ across markets, the number of observations varies across indices.

\paragraph{Construction and scaling of daily returns.}
Let $P_n$ denote the daily closing price and define the log-price as
$Y_n=\log P_n$. The raw log-price increment is
$Y_n-Y_{n-1}$. In the empirical analysis, returns are expressed in
percentage points and are therefore defined as
\begin{equation}
    r_n
    =
    100\left(Y_n-Y_{n-1}\right).
\end{equation}

Accordingly, since
\(
H_{n+1}=\int_n^{n+1}s_t\,dt
\)
is measured in squared percentage-return units, while $Y_t$ itself
remains on the raw log-price scale, we specify the continuous-time
return process as
\begin{equation}
    dY_t
    =
    0.01\sqrt{s_t}\,dW_t^Y,
\end{equation}
where $0.01=1/100$ converts the percentage-return scale of
$\sqrt{s_t}$ to the log-price scale.

Apart from this scaling of returns, all other empirical inputs,
including RV, are used on their original raw scale
without additional rescaling. The observed return $r_n$ therefore
corresponds to the percentage-scaled increment of the continuous-time
log-price process over one trading day.

\subsection{Benchmark Models Selection}
\label{sec:benchmark models}

We compare the proposed model with five established volatility benchmarks:
GARCH, Realized GARCH, HEAVY-r, and GARCH-It\^{o}, and HAR-QML. These models represent complementary approaches to volatility forecasting, differing primarily in
their use of realized-measure information and in whether volatility dynamics
are formulated in discrete or continuous time. Together, they provide a
structured basis for assessing the predictive value of realized-measure
information and continuous-time modeling in the proposed framework. To ensure comparability, all models are trained and evaluated under the same rolling-window framework, using the same data splits, refitting schedule, and forecast-evaluation protocol, while retaining their model-specific estimation procedures.
Throughout this subsection, notation follows Appendix~\ref{Appendix:B}.

\paragraph{GARCH.}
As the conventional return-based discrete-time benchmark, we use GARCH$(1,1)$ \citep{Bollerslev1986}. Using the same return equation as in Appendix~\ref{Appendix:B}, its conditional variance follows
\begin{equation}
    h_n
    =
    \omega
    + \alpha r_{n-1}^{2}
    + \beta h_{n-1}.
\end{equation}
GARCH serves as a standard low-frequency benchmark for assessing the additional forecasting value of realized-measure information and the continuous-time structure of the proposed model.

\paragraph{Realized GARCH.}
As the principal discrete-time realized-measure benchmark, we use the
log-linear Realized GARCH$(1,1)$ model \citep{Hansen2012} described in Appendix~\ref{Appendix:B}. Unlike the proposed framework, Realized GARCH
incorporates realized-measure information through discrete-time variance
updating and a measurement equation. The comparison assesses predictive
performance relative to an established measurement-driven volatility model.

\paragraph{HEAVY-r.}
As a high-frequency-information benchmark, we use the HEAVY-r$(1,1)$ model
\citep{Shephard2010},
\begin{equation}
    h_{n+1}
    =
    \omega+\alpha R_n+\beta h_n,
\end{equation}
HEAVY-r updates
conditional variance directly using realized-measure information, providing a parsimonious discrete-time benchmark for assessing the additional value of
the proposed continuous-time state dynamics.

\paragraph{GARCH-It\^o.}
As the continuous-time benchmark, we use GARCH-It\^o
\citep{KimWang2016}, with log-price dynamics
\begin{equation}
    dY_t=\sqrt{s_t}\,dW_t^Y,
\end{equation}
and conditional daily integrated variance
\begin{align}
    h_n
    &=
    \mathbb{E}
    \left[
        \int_{n-1}^{n}s_t\,dt
        \,\middle|\,
        \mathcal{F}_{n-1}
    \right], \\
    h_n
    &=
    \omega^{g}
    +\gamma h_{n-1}
    +\beta^{g}r_{n-1}^{2}.
\end{align}
GARCH-It\^o provides a continuous-time comparison that uses realized-variance
information in estimation, while the proposed model additionally introduces
a dynamic realized-measure state that feeds into latent variance. The
 comparison therefore assesses the predictive value of this
measurement-driven state structure beyond an established continuous-time
GARCH model.

\paragraph{HAR-QML}
As a realized-measure-based discrete-time benchmark, we use the heterogeneous autoregressive model estimated by quasi-maximum likelihood (HAR-QML), following the HAR structure of \citet{Corsi2009}. The one-day-ahead conditional variance forecast is specified as
\begin{equation}
    h_n
    =
    \beta_0
    + \beta_d RV_{n-1}
    + \beta_w RV_{n-1}^{(w)}
    + \beta_m RV_{n-1}^{(m)},
\end{equation}
where
\begin{equation}
    RV_{n-1}^{(w)}
    =
    \frac{1}{5}\sum_{j=1}^{5} RV_{n-j},
    \qquad
    RV_{n-1}^{(m)}
    =
    \frac{1}{22}\sum_{j=1}^{22} RV_{n-j},
\end{equation}
representing weekly and monthly realized-volatility components,
respectively. The conventional HAR-RV model is commonly estimated by ordinary least
squares (OLS) using realized variance as the dependent variable \citep{Corsi2009}. In our
benchmark comparison, however, we estimate the HAR specification using
a Gaussian quasi-likelihood for daily returns to improve comparability across all models. 

\subsection{Implementation of Correlated Brownian Innovations}
\label{app:correlated_brownian}

To implement the dependence structure in
Equation~\eqref{eq:brownian_corr}, the correlation parameters are first
constrained to lie within the admissible region specified in
Section~\ref{sec:ct_dependence}. We parameterize
\begin{align}
    \rho_{Y\!x}
    &=
    -m\,\operatorname{sigmoid}(\psi_x),
    \\
    \rho_{Y\!z}
    &=
    -m
    \sqrt{1-\rho_{Y\!x}^{2}}\,
    \operatorname{sigmoid}(\psi_z),
\end{align}
where $\psi_x,\psi_z\in\mathbb{R}$ are unconstrained parameters and
$m=0.995$ provide a small numerical margin. This parameterization keeps the
correlation structure strictly within the admissible region throughout estimation. We restrict both return-linked correlations to be non-positive, reflecting the conventional leverage direction, while maintaining the positive-semidefinite correlation constraint.

Given $\rho_{Y\!x}$ and $\rho_{Y\!z}$, the correlated Brownian innovations are
constructed from independent Brownian drivers through a linear
transformation, following standard simulation methods for correlated
diffusions \citep{KloedenPlaten1992,Glasserman2004}. Let
$\bar W_t^1$, $\bar W_t^2$, and $\bar W_t^3$ be mutually independent standard
Brownian motions, also independent of $U_t$. We define
\begin{align}
    dW_t^Y
    &=
    d\bar W_t^1,
    \\
    dW_t^x
    &=
    \rho_{Y\!x}\,d\bar W_t^1
    +
    \sqrt{1-\rho_{Y\!x}^{2}}\,d\bar W_t^2,
    \\
    dW_t^z
    &=
    \rho_{Y\!z}\,d\bar W_t^1
    +
    a_z\,d\bar W_t^2
    +
    b_z\,d\bar W_t^3,
\end{align}
where
\begin{align}
    a_z
    &=
    -\frac{\rho_{Y\!x}\rho_{Y\!z}}
    {\sqrt{1-\rho_{Y\!x}^{2}}},
    \\
    b_z
    &=
    \sqrt{
        1-\rho_{Y\!z}^{2}-a_z^{2}
    }.
\end{align}

The loading $a_z$ is chosen so that the latent-variance and realized-measure
innovations remain contemporaneously uncorrelated, while $b_z$ normalizes
$W_t^z$ to unit variance. Consequently,
\begin{align}
    d\langle W^Y,W^x\rangle_t
    &=
    \rho_{Y\!x}\,dt,
    \\
    d\langle W^Y,W^z\rangle_t
    &=
    \rho_{Y\!z}\,dt,
    \\
    d\langle W^x,W^z\rangle_t
    &=
    0,
\end{align}
with $U_t$ independent of the remaining innovations. This construction
therefore implements the dependence structure specified in
Section~\ref{sec:ct_dependence}.

\subsection{Estimation and Implementation Details}
\label{app:rolling_optimization}

\paragraph{Rolling-window protocol.}
For each index $i$, let $N_i$ denote the total number of daily
observations after data cleaning. We adopt a 50:50 hold-out design,
consistent with previous volatility-forecasting applications that use
the first half of the sample for estimation and the second half for
out-of-sample evaluation
\citep{balaban2006forecasting}.
More precisely, the fixed estimation-window length is
\begin{equation}
    W_i=\left\lfloor \frac{N_i}{2}\right\rfloor.
\end{equation}
so that the first $W_i$ observations form the initial in-sample estimation window and the remaining $N_i-W_i$ observations constitute the out-of-sample forecasting period. The estimation-window length is
kept fixed throughout the exercise rather than expanded over time.

Parameters are re-estimated every 21 trading days, approximately once
per month, following common practice in rolling volatility-forecasting
applications \citep{conrad2025mf2garch}. One-day-ahead forecasts are
generated on every trading day. Using one-based indexing, the initial
model is estimated from observations $1,\ldots,W_i$ and is held fixed
when forecasting $W_i+1,\ldots,W_i+21$. At the next re-estimation date,
the fixed-length window is advanced by 21 observations, so that the
model is re-estimated using observations $22,\ldots,W_i+21$, after
which forecasts are generated for $W_i+22,\ldots,W_i+42$.

More generally, at the $k$th re-estimation, $k=0,1,2,\ldots$, the
estimation sample is
\begin{equation}
    \{21k+1,\ldots,W_i+21k\},
\end{equation}
and the resulting parameter estimates are held fixed for the following
21 daily one-step-ahead forecasts, with the final block truncated at
the end of the sample. For each forecast target $t$, the forecast
origin is $t-1$, and only information available up to that origin is
used.

\paragraph{Optimization.}
Model parameters are estimated using the Adam stochastic optimization
algorithm \citep{kingma2015adam}, with a learning rate of $10^{-4}$
and mini-batches of 32 forecast origins. The initial in-sample fit is
optimized for 1,000 epochs, while each subsequent rolling
re-estimation is run for 300 epochs. We do not employ
validation-based early stopping. Instead, each estimation run is
completed for the prespecified number of epochs, and the parameter
configuration attaining the lowest in-sample training loss within that
run is retained.

At each rolling re-estimation date, the parameter vector is
initialized from the estimates obtained at the preceding fit. This
warm-start strategy exploits the temporal proximity of successive
rolling estimation windows while allowing the model to be fully
re-optimized using only observations contained in the current window.
The Adam optimizer state is re-initialized at each refit rather than
carried forward across estimation windows. Between consecutive
re-estimation dates, the estimated SDE parameters are held fixed and
used to generate daily one-step-ahead forecasts.

\paragraph{Simulation and numerical settings.}
The continuous-time state system is simulated using the
Euler-Maruyama discretization \citep{higham2001algorithmic} as
implemented in the \texttt{torchsde} framework
\citep{kidger2021efficient}. During parameter estimation, each
one-day transition is approximated using 10 equally spaced time steps,
corresponding to a discretization interval of $\Delta t=0.1$, and 64
Monte Carlo particles are used to approximate the simulation-based
training objective. For out-of-sample forecasting, numerical accuracy
is increased by using 20 time steps per trading day
($\Delta t=0.05$) and 256 Monte Carlo particles for each forecast
origin. These settings are held fixed across all indices and rolling
windows.

For each simulated particle, the one-day integrated variance
$ H_{n+1} = \int_n^{n+1}s_t\,dt$
is approximated numerically from the simulated latent variance path
using the trapezoidal rule
\citep{zheng2023multilevel}. The reported variance forecast is obtained
by averaging the particle-specific integrated variances across the
Monte Carlo ensemble. All
experiments use a fixed random seed of 1313 to improve computational reproducibility.

\subsection{Evaluation Metrics}
\label{sec:forecast_metrics}

Following standard volatility-forecast evaluation practice
\citep{HansenLunde2005,Patton2011}, we evaluate the one-day-ahead
integrated-variance forecasts using QLIKE, MSE, and MAE on the variance
scale. For  $T$ out-of-sample observations, let $\widehat H_n$ denote the
forecast of daily integrated variance and $R_n$ the corresponding observed
realized measure.

\paragraph{QLIKE.}
The Quasi-Likelihood (QLIKE) loss is

\begin{equation}
    \mathrm{QLIKE}
    =
    \frac{1}{T}
    \sum_{n=1}^{T}
    \left(
        \log \widehat H_n
        +
        \frac{R_n}{\widehat H_n}
    \right).
\end{equation}
QLIKE is widely used for volatility forecast evaluation and is robust to
noise in the realized-volatility proxy under appropriate conditions
\citep{Patton2011}. Lower values indicate better forecasting performance.

\paragraph{MSE.}
The variance-scale mean squared error is
\begin{equation}
    \mathrm{MSE}_{\mathrm{var}}
    =
    \frac{1}{T}
    \sum_{n=1}^{T}
    \left(
        R_n-\widehat H_n
    \right)^2.
\end{equation}
MSE places greater weight on large forecast errors.

\paragraph{MAE.}
The variance-scale mean absolute error is
\begin{equation}
    \mathrm{MAE}_{\mathrm{var}}
    =
    \frac{1}{T}
    \sum_{n=1}^{T}
    \left|
        R_n-\widehat H_n
    \right|.
\end{equation}
MAE penalizes forecast errors linearly and is therefore less sensitive to
large individual errors than MSE. Lower MSE and MAE values indicate better
forecasting performance.

\subsection{Statistical Forecast Comparison}
\label{app:statistical_comparison}

\paragraph{DM tests.}
Pairwise differences in predictive accuracy are assessed using the
DM test \citep{DieboldMariano1995}. For models $i$ and
$j$, let $L_n^{(i)}$ and $L_n^{(j)}$ denote their losses at forecast origin
$n$. The loss differential is
\begin{equation}
    d_n^{(i,j)}
    =
    L_n^{(i)}-L_n^{(j)},
\end{equation}
with sample mean
\begin{equation}
    \bar d^{(i,j)}
    =
    \frac{1}{T}\sum_{n=1}^{T}d_n^{(i,j)}.
\end{equation}
Thus, $\bar d^{(i,j)}<0$ indicates a lower average loss for model $i$, while
$\bar d^{(i,j)}>0$ favors model $j$. The null hypothesis of equal predictive
accuracy is
\begin{equation}
    H_0:
    \mathbb{E}\!\left[d_n^{(i,j)}\right]=0.
\end{equation}

The DM statistic is based on the mean loss differential relative to a
heteroskedasticity- and autocorrelation-consistent (HAC) estimate of its
long-run variance, allowing the sequence of forecast-loss differences to be
serially correlated \citep{DieboldMariano1995,NeweyWest1987}. We apply the
finite-sample correction of \citet{HarveyLeybourneNewbold1997} and report
two-sided $p$-values.

The HAC truncation lag is selected using the rule
\begin{equation}
    \ell
    =
    \left\lfloor
        4\left(\frac{T}{100}\right)^{2/9}
    \right\rfloor,
\end{equation}
following the commonly used Newey-West bandwidth rule
\citep{NeweyWest1994}. Statistical significance therefore indicates evidence
against equal predictive accuracy, with the sign of $\bar d^{(i,j)}$
determining which model achieves the lower loss.

\paragraph{MCS tests.}
To account for multiple-model comparisons, we apply the Model Confidence Set
procedure of \citet{Hansen2011} separately for each equity index
and loss criterion. Starting from the full set of candidate specifications,
the procedure sequentially eliminates models with inferior predictive
performance and reports the resulting 90\% confidence set
($\alpha=0.10$).

Inference is based on $B=5{,}000$ stationary-bootstrap replications
\citep{PolitisRomano1994}, which account for temporal dependence in the out-of-sample loss differentials by resampling blocks of consecutive observations. The expected bootstrap block length is set to

\begin{equation}
    w = 1.5T^{1/3},
\end{equation}
where $T$ denotes the number of aligned out-of-sample observations. We use
the range statistic $T_R$ of \citet{Hansen2011} to test equal
predictive ability and identify the model to be eliminated at each step.
Elimination continues until equal predictive ability can no longer be
rejected among the surviving models. The final set therefore contains the
specifications for which the data do not provide sufficient evidence of
inferior predictive performance under the corresponding loss criterion.

\subsection{Detailed Results for Ablation Study}
\label{app:ablation_results}
\begin{table*}[htbp]
\centering
\caption{DM Comparison of the Full SDE and the No-Feedback Specification}
\label{tab:dm_full_vs_gamma0}
\small
\setlength{\tabcolsep}{5pt}
\renewcommand{\arraystretch}{1.08}

\begin{tabular}{lcccccc}
\toprule
& \multicolumn{2}{c}{QLIKE}
& \multicolumn{2}{c}{MSE$_{\mathrm{var}}$}
& \multicolumn{2}{c}{MAE$_{\mathrm{var}}$} \\
\cmidrule(lr){2-3}
\cmidrule(lr){4-5}
\cmidrule(lr){6-7}
Index
& Full & $\gamma=0$
& Full & $\gamma=0$
& Full & $\gamma=0$ \\
\midrule
SPX
& \textbf{0.0709}$^\dagger$ & 0.0778
& \textbf{2.6538} & 2.6836
& \textbf{0.4285}$^\dagger$ & 0.4333 \\

SSMI
& \textbf{0.1724}$^\dagger$ & 0.1779
& \textbf{3.5165} & 3.5643
& \textbf{0.3812}$^\dagger$ & 0.3890 \\

STI
& \textbf{-0.1208}$^\dagger$ & -0.1145
& \textbf{0.2271}$^\dagger$ & 0.2329
& \textbf{0.2367}$^\dagger$ & 0.2426 \\

OMXSPI
& \textbf{0.0941}$^\dagger$ & 0.0998
& \textbf{0.4805} & 0.4905
& \textbf{0.3013}$^\dagger$ & 0.3061 \\

KS11
& \textbf{0.0713}$^\dagger$ & 0.0753
& \textbf{0.5928} & 0.5940
& \textbf{0.2840}$^\dagger$ & 0.2882 \\

OMXHPI
& \textbf{0.1767}$^\dagger$ & 0.1805
& \textbf{0.3433}$^\dagger$ & 0.3471
& \textbf{0.3106}$^\dagger$ & 0.3149 \\

SSEC
& \textbf{0.7730}$^\dagger$ & 0.7760
& \textbf{4.4282} & 4.4865
& \textbf{0.6749}$^\dagger$ & 0.6778 \\

KSE
& \textbf{0.3161} & 0.3166
& \textbf{1.0611} & 1.0618
& \textbf{0.3866}$^\dagger$ & 0.3885 \\

MXX
& \textbf{0.4577} & 0.4589
& 1.8498 & \textbf{1.8415}$^\dagger$
& \textbf{0.4261}$^\dagger$ & 0.4272 \\

IXIC
& \textbf{0.2686}$^\dagger$ & 0.2735
& \textbf{2.2612} & 2.3074
& \textbf{0.4341}$^\dagger$ & 0.4386 \\

OMXC20
& \textbf{0.7772} & 0.7773
& 5.1722 & \textbf{5.1273}
& \textbf{0.5813} & 0.5814 \\

STOXX50E
& 0.8867 & \textbf{0.8606}
& \textbf{4.0608} & 4.1081
& \textbf{0.5899} & 0.5907 \\

SMSI
& \textbf{0.7595}$^\dagger$ & 0.7615
& \textbf{4.6071} & 4.6193
& \textbf{0.5146} & 0.5161 \\

N225
& \textbf{0.4498}$^\dagger$ & 0.4526
& \textbf{1.9749} & 1.9794
& \textbf{0.4754}$^\dagger$ & 0.4774 \\

OSEAX
& 0.6341 & \textbf{0.6340}
& 20.6389 & \textbf{20.4979}
& \textbf{0.6978} & 0.6982 \\

RUT
& \textbf{0.3307}$^\dagger$ & 0.3330
& \textbf{1.3142} & 1.3351
& \textbf{0.4272}$^\dagger$ & 0.4300 \\

NSEI
& 0.4990 & \textbf{0.4903}
& 5.0865 & \textbf{5.0525}
& \textbf{0.4379}$^\dagger$ & 0.4412 \\

\midrule
Lower mean loss
& \textbf{14} & 3
& \textbf{13} & 4
& \textbf{17} & 0 \\

Significantly better (10\%)
& \textbf{11} & 0
& \textbf{2} & 1
& \textbf{13} & 0 \\
\bottomrule
\end{tabular}
\vspace{0.3em}
\begin{minipage}{0.95\linewidth}
\footnotesize
\textit{Notes:}
\textit{Notes:}
Full denotes the proposed SDE with realized-measure feedback, while
$\gamma=0$ removes the feedback term from the latent log-variance drift.
Lower losses indicate better forecasting performance. Bold entries denote
the lower mean loss within each pair. A dagger indicates that the lower-loss
model significantly outperforms its counterpart under the two-sided DM test
at the 10\% level ($^{\dagger}p<0.10$). The final two rows report the number
of indices for which each specification achieves the lower mean loss and a
significant improvement, respectively.
\end{minipage}
\end{table*}

\begin{table*}[htbp]
\centering
\caption{DM Comparison of the Full SDE and the
Independent-Brownian Specification}
\label{tab:dm_full_vs_independent}
\small
\setlength{\tabcolsep}{5pt}
\renewcommand{\arraystretch}{1.08}

\begin{tabular}{lcccccc}
\toprule
& \multicolumn{2}{c}{QLIKE}
& \multicolumn{2}{c}{MSE$_{\mathrm{var}}$}
& \multicolumn{2}{c}{MAE$_{\mathrm{var}}$} \\
\cmidrule(lr){2-3}
\cmidrule(lr){4-5}
\cmidrule(lr){6-7}
Index
& Full & Indep.
& Full & Indep.
& Full & Indep. \\
\midrule
SPX
& 0.0709 & \textbf{0.0703}
& 2.6538 & \textbf{2.6150}
& 0.4285 & \textbf{0.4269} \\

SSMI
& 0.1724 & \textbf{0.1720}
& \textbf{3.5165} & 3.5314
& \textbf{0.3812} & 0.3813 \\

STI
& -0.1208 & \textbf{-0.1215}
& \textbf{0.2271} & 0.2280
& 0.2367 & \textbf{0.2361} \\

OMXSPI
& \textbf{0.0941}$^\dagger$ & 0.0952
& \textbf{0.4805} & 0.4818
& \textbf{0.3013} & 0.3017 \\

KS11
& 0.0713 & \textbf{0.0712}
& 0.5928 & \textbf{0.5914}
& 0.2840 & \textbf{0.2833} \\

OMXHPI
& \textbf{0.1767}$^\dagger$ & 0.1773
& \textbf{0.3433} & 0.3443
& \textbf{0.3106}$^\dagger$ & 0.3121 \\

SSEC
& \textbf{0.7730} & 0.7734
& \textbf{4.4282} & 4.4310
& \textbf{0.6749} & 0.6757 \\

KSE
& \textbf{0.3161} & 0.3162
& 1.0611 & \textbf{1.0595}
& \textbf{0.3866} & 0.3867 \\

MXX
& 0.4577 & \textbf{0.4572}
& \textbf{1.8498} & 1.8669
& \textbf{0.4261} & 0.4269 \\

IXIC
& \textbf{0.2686} & 0.2691
& \textbf{2.2612} & 2.2747
& \textbf{0.4341} & 0.4344 \\

OMXC20
& 0.7772 & \textbf{0.7762}
& \textbf{5.1722} & 5.1755
& \textbf{0.5813} & 0.5818 \\

STOXX50E
& 0.8867 & \textbf{0.8850}
& \textbf{4.0608} & 4.0931
& \textbf{0.5899} & 0.5903 \\

SMSI
& \textbf{0.7595} & 0.7597
& \textbf{4.6071} & 4.6205
& \textbf{0.5146} & 0.5159 \\

N225
& \textbf{0.4498}$^\dagger$ & 0.4523
& \textbf{1.9749} & 1.9806
& \textbf{0.4754} & 0.4753 \\

OSEAX
& \textbf{0.6341} & 0.6347
& \textbf{20.6389} & 20.7068
& 0.6978 & \textbf{0.6976} \\

RUT
& \textbf{0.3307} & 0.3309
& 1.3142 & \textbf{1.3019}
& 0.4272 & \textbf{0.4264} \\

NSEI
& \textbf{0.4990} & 0.4995
& 5.0865 & \textbf{5.0603}
& 0.4379 & \textbf{0.4377} \\

\midrule
Lower mean loss
& \textbf{10} & 7
& \textbf{12} & 5
& \textbf{11} & 6 \\

Significantly better (10\%)
& \textbf{3} & 0
& 0 & 0
& \textbf{1} & 0 \\
\bottomrule
\end{tabular}
\vspace{0.3em}
\begin{minipage}{0.95\linewidth}
\footnotesize
\textit{Notes:}
Full denotes the proposed SDE with return-linked Brownian dependence, while
Indep.\ imposes $\rho_{Yx}=\rho_{Yz}=0$. Boldface and dagger notation follow
Table~\ref{tab:dm_full_vs_gamma0}. 
\end{minipage}

\end{table*}

\FloatBarrier

\subsection{Sensitivity Analysis}
\label{app:sensitivity}

\begin{table}[htbp]
\centering
\caption{Sensitivity Analysis}
\label{tab:sensitivity}
\small
\begin{tabular}{llccc}
\toprule
Sensitivity 
& Configurations 
& Avg. $\Delta$QLIKE 
& Avg. $\Delta$MSE$_{\mathrm{var}}$
& Avg. $\Delta$MAE$_{\mathrm{var}}$ \\
\midrule

Random seed
    & 7, 42, 2024
    & 0.0012
    & 0.0178
    & 0.0011 \\

Training simulation paths
    & 16, 32, 128
    & 0.0021
    & 0.0147
    & 0.0009 \\

Forecast simulation paths
    & 64, 128, 256
    & 0.0017
    & 0.0215
    & 0.0010 \\

Training discretization
    & 5, 10, 20 substeps
    & 0.0010
    & 0.0128
    & 0.0008 \\

\bottomrule
\end{tabular}

\vspace{0.4em}
\begin{minipage}{0.92\linewidth}
\footnotesize
\textit{Notes:}
Sensitivity is assessed using QLIKE, MSE$_{\mathrm{var}}$, and
MAE$_{\mathrm{var}}$ across the 17 indices.
For each metric, $\Delta$ denotes the difference between the maximum and
minimum loss across the reported configurations for each index, averaged
across the 17 indices.
The small average variations indicate that forecasting performance is
stable with respect to the reported computational configurations.
\end{minipage}
\end{table}

\end{document}